\documentclass[11pt]{article}
\usepackage{osid}
\usepackage{graphicx}
\usepackage{latexsym}
\usepackage{xypic}
\usepackage{amssymb}
\usepackage{multirow}
\usepackage{amsmath}
\usepackage{amscd}
\usepackage{mathrsfs}
\usepackage{bm}
\usepackage{theorem}
\usepackage{epsf}

\renewcommand{\RHD}{Gresnigt {\em et.\ al.}, The Stabilized Poincare-Heisenberg algebra: a Clifford algebra viewpoint}
\renewcommand{\LHD}{Gresnigt {\em et.\ al.}, The Stabilized Poincare-Heisenberg algebra: a Clifford algebra viewpoint}

\begin{document}

\title{The Stabilized Poincare-Heisenberg algebra: a Clifford algebra viewpoint}
\author{ N. G. Gresnigt$^*$, P. F. Renaud$^\dagger$, P. H. Butler$^*$ \medskip\\
{\scriptsize \noindent $^*$Department of Physics and Astronomy, University of Canterbury, Private Bag 4800, Christchurch, New Zealand\\
\noindent $^\dagger$Department of Mathematics and Statistics,
University of Canterbury, Private Bag 4800, Christchurch, New
Zealand\\} }

\maketitle

\begin{abstract}
The stabilized Poincare-Heisenberg algebra (SPHA) is the Lie algebra
of quantum relativistic kinematics generated by fifteen generators.
It is obtained from imposing stability conditions after attempting
to combine the Lie algebras of quantum mechanics and relativity
which by themselves are stable, however not when combined. In this
paper we show how the sixteen dimensional Clifford algebra
$C\ell(1,3)$ can be used to generate the SPHA. The Clifford algebra
path to the SPHA avoids the traditional stability considerations,
relying instead on the fact that $C\ell(1,3)$ is a semi-simple algebra and therefore stable. 
It is therefore conceptually easier and more straightforward to work with a Clifford algebra. 
The Clifford algebra path suggests the next evolutionary step toward a
theory of physics at the interface of GR and QM might be to depart from working in space-time and
instead to work in space-time-momentum.
\end{abstract}

\section{Introduction}                                               %
Physical theories are merely approximations to the natural world and
the physical constants involved cannot be known without some degree
of uncertainty. Properties of a model that are sensitive to
small changes in the model, in particular changes in the values of the
parameters, are unlikely to be observed. It can thus be reasoned
that one should search for physical theories which do not change in
a qualitative matter under a small change of the parameters. Such
theories are said to be physically {\it stable}. 

This concept of the physical stability of a theory can be given a mathematical meaning as follows.
A mathematical structure is said to be mathematically stable for a class of
deformations if  any deformation in this class leads to an
isomorphic structure. More precisely, a Lie algebra is said to be
stable if small perturbations in its structure constants lead to
isomorphic Lie algebras. The idea of mathematical stability provides insight into
the validity of a physical theory or the need for a generalization of the
theory. If a theory is not stable, one might choose to deform it
until a stable theory is reached. Such a stable theory is likely to
be a generalization of wider validity compared to the original
unstable theory.

Lie algebraic deformation theory has been historically successful. Snyder \cite{Snyder:1946qz} in 1947 showed that the assumption that spacetime be a continuum is not required for Lorentz invariance. Snyder's framework however leads to a lack of translational invariance, which later in the same year, Yang \cite{Yang:1947ud} showed can be corrected if one allows for spacetime to be curved. Yang, in the same paper also presented the complete Lie algebra associated with the suggested corrections. It was Mendes \cite{VilelaMendes:1994zg} who in the last decade concluded that when one considers the Poincare and Heisenberg algebras together, the resultant Poincare-Heisenberg algebra is not a stable Lie algebra. Mendes showed however that the algebra can be stabilized, requiring two additional length scales. The stabilized algebra is the same as the algebra obtained by Yang in 1947. It was Faddeev \cite{Faddeev} and Mendes who noted that, in hindsight, stability considerations could have predicted the relativistic and quantum revolutions of the last century. Chryssomalakos and Okon \cite{Chryssomalakos:2004gk,Chryssomalakos:2004wc} showed that by a suitable identification of the generators, triply special relativity proposed by Kowalski-Glikman and Smolin \cite{Kowalski-Glikman:2004kp} can be brought to a linear form and that the resulting algebra is again the same as Yang's algebra. 

More recently, stability considerations have led to the  stabilized
Poincare-Heisenberg algebra (SPHA) as the favorite candidate for the
Lie algebra describing physics at the interface of GR and
QM. Chryssomalakos and Okon \cite{Chryssomalakos:2004gk} showed uniqueness of the SPHA. Incorporating gravitational effects in
quantum measurement of spacetime events renders spacetime
non-commutative and leads to modifications in the fundamental
commutators \cite{Ahluwalia:1993dd}. In 2005, Ahluwalia-Khalilova \cite{Ahluwalia-Khalilova:2005jn,\cite{Ahluwalia-Khalilova:2005km}} showed that the fact that the Heisenberg fundamental commutator, $[X_\mu,P_\mu]=i\hbar$, undergoes non-trivial modifications at the interface of GR and QM, suggests quantum mechanics and relativity will undergo numerous corrections and modification including modification of the position-momentum Heisenberg uncertainty relations.
On top of this, spacetime in SPHA requires two new
length scales, one in the extreme short distance range, the other on
the cosmological scale. We denote these new length scales $\ell_p$
and $\ell_c$ respectively (following the notation used in
\cite{Ahluwalia-Khalilova:2005jn}).

In addition, SPHA points to the existence of another dimensionless
constant $\beta$ (or $\alpha_3$ in \cite{Chryssomalakos:2004gk}) which, if nonzero,
will radically affect some of the quantum relativistic notions. The
presence of $\beta$ has been noted in \cite{VilelaMendes:1994zg,Chryssomalakos:2004gk,Khruschev:2002cq}
with different emphasis. Chryssomalakos and Okon \cite{Chryssomalakos:2004gk} show
that it is always possible to gauge away this dimensionless constant by a
suitable redefinition of the generators. The overall meaning of
$\beta$ seems to not be well understood in the literature. We show that Clifford algebra allows one to obtain an explicit expression for $\beta$ in terms of an angle parameter $\varphi$. This is a step toward
understanding how $\beta$ will affect various quantum relativistic
notions.

Our approach to the SPHA is quite  different in that we adopt a
Clifford algebra perspective. A physical theory should have its
roots in the geometry underlying the theory and be represented in a
way that captures this geometry accordingly. Clifford (geometric)
algebra is the appropriate tool that allows us to naturally
encompass the underlying geometry of the space we are working in.
Although a brief overview of Clifford algebra will be presented in
section 3 of this paper, we omit any  in-depth discussion of
Clifford algebra and how it arises from geometry. The reader is
instead referred to the following texts by Lounesto \cite{pl},
Hestenes \cite{dh1,dh5} and Doran and Lasenby \cite{dl}.

The SPHA has fifteen generators. We wish to represent these
generators by elements of the Clifford algebra $C\ell(1,3)$ which is
a sixteen dimensional algebra. The non-scalar basis elements of
$C\ell(1,3)$  can be used to generate the SPHA by taking commutators
 with the Clifford product. ($C\ell(1,3)$ is chosen
over $C\ell(3,1)$ or other sixteen dimensional Clifford algebras for
reasons explained in \cite{wjpb,pblm}.) It is interesting to note
that using this Clifford algebra, one avoids all stability
considerations. This is because the Clifford algebra $C\ell(1,3)$ is semi-simple and
therefore stable. Some further discussion on this is reserved for the next
section. Ahluwalia-Khalilova \cite{Ahluwalia:1993dd} noted that incorporating gravitational effects
in quantum measurement of spacetime events renders spacetime
non-commutative and leads to modifications in the fundamental
commutators. This non-commutativity of space-time and modifications
to the fundamental commutators arise naturally from Clifford
algebra. There thus seems to be a number of reasons why one should
consider using Clifford algebra in theories of quantum gravity.

Since we are logically led to consider two additional length scales as well as a dimensionless
constant in the proposed new algebra for kinematics at the interface of GR and QM, it is important to note that on heuristic grounds Amelino-Camelia \cite{Amelino-Camelia:2003bt} and Smolin \cite{Kowalski-Glikman:2004kp} have also considered such a path. However, in the work presented here such resulting modifications to the quantum and spacetime structure arise
not as an ad hoc speculation but from the Principle of Lie Algebraic
stability. The reader is referred to \cite{Judes:2002bw,Grumiller:2003df,Schutzhold:2003yp,Chryssomalakos:2004gk,Ahluwalia-Khalilova:2005jn,Christian:2006ut}
for additional discussion of these and related issues.

\section{Stability Theory}                                           %
As noted by Mendes \cite{VilelaMendes:1994zg} and Faddeev \cite{Faddeev}, in hindsight, 
the paradigm of algebraic deformations to obtain
stable theories  has the power to have predicted the relativistic
and the quantum revolutions of the last century. This section will
show how algebraic deformations lead from the Poisson algebra of
classical mechanics to the Heisenberg algebra of quantum mechanics
and from the Galilean algebra of Galilean relativity to the Poincare
algebra of special relativity. The theory of Lie-algebraic
deformations is not discussed in detail here. For a thorough and complete
treatment the reader is referred to Gerstenhaber
\cite{Gerstenhaber}, Nijenhuis and Richardson \cite{NandR} and
Chryssomalakos and Okon \cite{Chryssomalakos:2004gk}.

When considering the Poisson and Galilean algebras, one finds that
the algebraic  structures are unstable. It is however possible to
stabilize both of the algebras. Doing so requires two deformation
parameters. These turn out to be the physical constants
$\frac{1}{c^2}$ and $\hbar$ for the Galilean algebra and Poisson
algebra respectively. Chryssomalakos and Okon \cite{Chryssomalakos:2004gk} explain that 
both the Galilean and Poisson algebra cases, the deformed algebras are 
isomorphic for all non-zero values of $\frac{1}{c^2}$ and $\hbar$. 
The values of these deformation parameters are determined by experiment.

Neither the Heisenberg nor the Poincare algebras preserve their
stability at the interface of GR and QM. A first attempt to find an algebra describing
physics at the interface of GR and QM may be to take the direct sum of the Poincare
and Heisenberg algebras to give the Poincare-Heisenberg algebra but
this algebra however is not stable. Mendes \cite{Chryssomalakos:2004gk} and
Chryssomalakos and Okon \cite{Chryssomalakos:2004gk} have both emphasised that one of
the important criteria to consider for a theory to be physically
viable, is the stability of the underlying Lie algebras, and so the
Poincare-Heisenberg algebra fails to be the algebra we desire.

The Poincare-Heisenberg algebra can however  be stabilized,
requiring two additional length scales and a dimensionless constant
in the same manner that the stabilization of the algebras of
Galilean and classical kinematics requires two constants
$\frac{1}{c^2}$ and $\hbar$. The resulting algebra is the stabilized
Poincare-Heisenberg algebra (SPHA), the algebra of kinematics at the
interface of GR and QM. It will be shown later that the SPHA is precisely the algebra
obtained by taking commutators of the elements of the Clifford space-time
algebra $C\ell(1,3)$.

\section{The Stabilised Poincare-Heisenberg Algebra}                 %
As noted above, the stabilised Poincare-Heisenberg algebra comes from combining
quantum mechanics and  relativity to get a stable theory of
kinematics at the interface of GR and QM. 
The algebra is given below in a similar form to that of
Chryssomalakos and Okon \cite{Chryssomalakos:2004gk} (which in turn is a somewhat more
abstract and mathematical version than the one given by
Ahluwalia-Khalilova \cite{Ahluwalia-Khalilova:2005jn} which focuses more on the
physical aspects of the algebra).

\begin{eqnarray}
\label{sph1}[iJ_{\mu\nu},iJ_{\rho\sigma}]=-(\eta_{\nu\rho}J_{\mu\sigma}+\eta_{\mu\sigma}J_{\nu\rho}-\eta_{\mu\rho}J_{\nu\sigma}-\eta_{\nu\sigma}J_{\mu\rho})
\end{eqnarray}
\begin{eqnarray}
\label{sph2}[iJ_{\mu\nu},P_{\lambda}]=-(\eta_{\nu\lambda}P_{\mu}-\eta_{\mu\lambda}P_{\nu})
\end{eqnarray}
\begin{eqnarray}
\label{sph3}[iJ_{\mu\nu},X_{\lambda}]=-\hbar(\eta_{\nu\lambda}X_{\mu}-\eta_{\mu\lambda}X_{\nu})
\end{eqnarray}
\begin{eqnarray}
\label{sph4}[P_{\mu},P_{\nu}]=q\alpha_1 J_{\mu\nu}
\end{eqnarray}
\begin{eqnarray}
\label{sph5}[X_{\mu},X_{\nu}]=q\alpha_2 J_{\mu\nu}
\end{eqnarray}
\begin{eqnarray}
\label{sph6}\label{PHPX}[P_{\mu},X_{\nu}]=q\eta_{\mu\nu}M+q\alpha_3 J_{\mu\nu}
\end{eqnarray}
\begin{eqnarray}
\label{PHPF}[P_{\mu},iM]=-\alpha_1 X_{\mu}+\alpha_3 P_{\mu}
\end{eqnarray}
\begin{eqnarray}
\label{PHXF}[X_{\mu},iM]=-\alpha_3 X_{\mu}+\alpha_2 P_{\mu}
\end{eqnarray}
\begin{eqnarray}
\label{sph9}[iJ_{\mu\nu},iM]=0
\end{eqnarray}

This paper is primarily concerned with finding a Clifford representation of the SPHA.
For this reason we focus on the mathematical theory and therefore adopt the notation used by Chryssomalakos and Okon \cite{Chryssomalakos:2004gk} instead of the notation used by Ahluwalia-Khalilova \cite{Ahluwalia-Khalilova:2005jn}, with the exception of the part of section 5.

\section{The Clifford Algebra $C\ell(1,3)$}                          %
The Clifford algebra $C\ell(1,3)$ is a 16  dimensional associative
algebra with a basis consisting of one scalar $1$, four
vectors $e_{\mu}$, six bivectors $e_{\mu\nu}$,
three trivectors $e_{\mu\nu\rho}$ and a pseudoscalar
$e_{\mu\nu\rho\sigma}=e_{0123}$ which for brevity
we write $e$, together with the Clifford product (geometric
product) used to multiply elements of the algebra. Duals of elements
of the Clifford algebra are found by multiplying through by the
pseudoscalar $e=e_{0123}$. Trivectors are thus
dual to vectors. Instead of writing trivectors as
$e_{\mu\nu\rho}$, we write them as
$ee_{\mu}$ for reasons which will become clear
later.

As noted earlier, we do not have to concern ourselves with the issue of stability
when using a Clifford algebra. All Clifford algebras are stable because the diagonal entries of the 
metric $\eta_{\mu\nu}$ are non-zero and therefore under a small perturbation,
the signature remains unchanged.

We are looking for a Clifford representation of the
stabilized Poincare-Heisenberg algebra. This means that
$X_{\mu}$, $P_{\mu}$, $iJ_{\mu\nu}$ and $iM$ are to be represented by
elements of a real Clifford algebra. The stabilized Poincare-Heisenberg
algebra is a Lie algebra so that we will require the Lie product to
be a commutator in the Clifford algebra, i.e. $[x,y]=xy-yx$.
The scalar element of the Clifford algebra commutes with every
element leaving us with 15 generators. We will start with $\omega_1 e_{\mu}$,
$\omega_2 ee_{\mu}$, $\omega_{3} e_{\mu\nu}$ and
$\omega_{4}e$, where $\omega_{i}\in\mathbb{R},
\;i=1,2,3,4$, and calculate their commutators. 

\begin{eqnarray}
\label{first}[\omega_{3}e_{\mu\nu},\omega_{3}e_{\rho\sigma}]&=&2\omega_{3}^2(\eta_{\nu\rho}e_{\mu\sigma}+\eta_{\mu\sigma}e_{\nu\rho}-\eta_{\mu\rho}e_{\nu\sigma}-\eta_{\nu\sigma}e_{\mu\rho})\\
\notag[\omega_3 e_{\mu\nu},\omega_2 ee_{\rho}]&=&2\omega_2 \omega_3(\eta_{\nu\rho}ee_{\mu}-\eta_{\mu\rho}ee_{\nu})\\
&=& 2\omega_3(\eta_{\nu\rho}(\omega_2 ee_{\mu})-\eta_{\mu\rho}(\omega_2 ee_{\nu}))\\
\notag[\omega_3 e_{\mu\nu},\omega_1 e_{\rho}]&=&2\omega_1 \omega_3(\eta_{\nu\rho}e_{\mu}-\eta_{\mu\rho}e_{\nu})\\
&=& 2\omega_2(\eta_{\nu\rho}(\omega_1 e_{\mu})-\eta_{\mu\rho}(\omega_1 e_{\nu}))
\end{eqnarray}

\begin{eqnarray}
\notag[\omega_2ee_{\mu},\omega_2ee_{\nu}]&=&2\omega_{2}^2e_{\mu\nu}\\
&=&2\frac{\omega_{2}^2}{\omega_{3}}(\omega_3e_{\mu\nu})\\
\notag[\omega_1e_{\mu},\omega_1e_{\nu}]&=&2\omega_{1}^2e_{\mu\nu}\\
&=&2\frac{\omega_{1}^2}{\omega_{3}}(\omega_3e_{\mu\nu})\\
\notag[\omega_2ee_{\mu}, \omega_1e_{\nu}]&=&2\omega_1\omega_2\eta_{\mu\nu}e\\
&=&2\eta_{\mu\nu}\frac{\omega_1\omega_2}{\omega_4}(\omega_4e)\\
\notag[\omega_2ee_{\mu},\omega_4e]&=&2\omega_2\omega_4e_{\mu}\\
&=&2\frac{\omega_2\omega_4}{\omega_1}(\omega_1e_{\mu})\\
\notag[\omega_1e_{\mu}, \omega_4e]&=&-2\omega_1\omega_4ee_{\mu}\\
&=&-2\frac{\omega_1\omega_4}{\omega_2}(\omega_2ee_{\mu})\\
\label{last}[\omega_3e_{\mu\nu},\omega_4e]&=&0
\end{eqnarray}

By defining the 15 operators
\begin{eqnarray}
X_{\mu}=\omega_1e_\mu=\frac{1}{2}\sqrt{-q\alpha_2}e_{\mu}
\end{eqnarray}
\begin{eqnarray}
P_{\mu}=\omega_2ee_{\mu}=\frac{1}{2}\sqrt{-q\alpha_1}ee_{\mu}
\end{eqnarray}
\begin{eqnarray}
iJ_{\mu\nu}=\omega_3e_{\mu\nu}=\frac{-1}{2}e_{\mu\nu}
\end{eqnarray}
\begin{eqnarray}
iM=\omega_4e=\frac{-1}{2}\sqrt{\alpha_1 \alpha_2}e
\end{eqnarray}
we obtain the stabilized
Poincare-Heisenberg algebra for the special case  where $\alpha_3$
is equal to zero. ( We discuss this point in the next section.) We call $X_{\mu}$, $P_{\mu}$, $iJ_{\mu\nu}$ and $iM$ above the Clifford generators of the stabilised Poincare-Heisenberg algebra. As in \cite{wjpb, pblm},
we interpret $X_\mu$ to be the position vectors, $P_\mu$ to be the
momenta and $iJ_{\mu\nu}$ to be the rotations and boosts.

\section{Clifford representation with $\alpha_3 \neq 0$}             %
In this section we will transform the Clifford representation in such a way
that the transformed  Clifford generators will generate the entire
stabilized Poincare-Heisenberg algebra rather than just the special
case where $\alpha_3$ is equal to zero. The physical interpretation
of this transformation will be discussed in the following section
and gives rise to a new concept in physics.

We start by defining by redefining $X_\mu$ as
\begin{eqnarray}
X_\mu=a\,e_\mu +b\,ee_\mu
\end{eqnarray}
giving
\[
X_\mu X_\nu=(a^2+b^2)e_{\mu\nu}
\]
and therefore
\begin{eqnarray}
\left[X_\mu,X_\nu\right]=2(a^2+b^2)\,e_{\mu\nu},\qquad \mu\neq\nu
\end{eqnarray}
What we require is;
\[
\left[X_\mu,X_\nu\right]=i\,q\,\alpha_2\,J_{\mu\nu}
\]
To get this we must thus define;
\begin{eqnarray}
iJ_{\mu\nu}= \frac{2(a^2+b^2)}{q\,\alpha_2}\,\,e_{\mu\nu},\qquad \mu\neq\nu
\end{eqnarray}
Similarly, define $P_{\mu}$ as
\begin{eqnarray}
P_\mu=d\,ee_\mu+c\,e_\mu
\end{eqnarray}
giving
\begin{eqnarray*}
P_\mu P_\nu=(c^2+d^2)e_{\mu\nu}
\end{eqnarray*}
and
\begin{eqnarray}
\left[P_\mu,P_\nu\right]=2(c^2+d^2)\,e_{\mu\nu},\qquad \mu\neq\nu
\end{eqnarray}
We want this to be equal to
\[
\left[P_\mu,P_\nu\right]=i\,q\,\alpha_1\,J_{\mu\nu}
\]
which requires
\begin{eqnarray}
iJ_{\mu\nu}= \frac{2(c^2+d^2)}{q\,\alpha_1}\,\,e_{\mu\nu},\qquad
\mu\neq\nu
\end{eqnarray}
Note that consistency in the definition for $J_{\mu\nu}$ implies
that
\[
\frac{a^2+b^2}{\alpha_2}=\frac{c^2+d^2}{\alpha_1}
\]

The above transformed definitions for $X_\mu$ and $P_\mu$ can be written in the form
\begin{eqnarray*}
\left[\begin{matrix}X_\mu\\P_\mu\end{matrix}\right]&=&
\left[\begin{matrix}a&b\\c&d\end{matrix}\right]\,
\left[\begin{matrix}e_\mu\\ee_\mu\end{matrix}\right]
\end{eqnarray*}
and
therefore
\begin{eqnarray*}
\left[\begin{matrix}e_\mu\\ee_\mu\end{matrix}\right]&=&\frac{1}{ad-bc}
\left[\begin{matrix}d&-b\\-c&a\end{matrix}\right]\,
\left[\begin{matrix}X_\mu\\P_\mu\end{matrix}\right]
\end{eqnarray*}
or
\begin{eqnarray}
e_\mu&=&\frac{1}{\Delta}(dX_\mu-bP_\mu)\\
ee_\mu&=&\frac{1}{\Delta}(-cX_\mu+aP_\mu)
\end{eqnarray}
where $\Delta=ad-bc$ is the determinant of the matrix.

Working out the commutator $[P_\mu,X_\nu]$ with the new transformed expressions for $X_\mu$ and $P_\mu$, we get
\begin{eqnarray*}
P_\mu X_\nu&=&(dee_\mu+ce_\mu)(ae_\nu+bee_\nu)\\
&=& (ac+bd)e_{\mu\nu}+(ad-bc)ee_{\mu\nu}
\end{eqnarray*}
while
\[
X_\nu P_\mu=(ac+bd)e_{\nu\mu}-(ad-bc)ee_{\nu\mu}
\]
So, if $\mu=\nu$,
\begin{eqnarray*}
\left[P_\mu,X_\nu\right]&=&P_\mu X_\mu-X_\mu P_\mu\\
&=&(ac+bd)\eta_{\mu\nu}+(ad-bc)\eta_{\mu\nu}\,e\\
&&-(ac+bd)\eta_{\mu\nu}+(ad-bc)\eta_{\mu\nu}\,e\\
&=&2(ad-bc)\eta_{\mu\nu}\,e
\end{eqnarray*}
while for $\mu\neq\nu$,
\begin{eqnarray*}
\left[P_\mu,X_\nu\right]=2(ac+bd)e_{\mu\nu}
\end{eqnarray*}
We want this to be equal to
\[
\left[P_\mu,X_\nu\right]=i\,q\eta_{\mu\nu}M+i\,q\alpha_3J_{\mu\nu}
\]
so that we have
\begin{eqnarray}
(\mu=\nu)\quad i\,q\, M=2(ad-bc)e
\end{eqnarray}
and
\begin{eqnarray}
(\mu\neq\nu)\quad i\,q\alpha_3J_{\mu\nu}=2(ac+bd)e_{\mu\nu}
\end{eqnarray}
i.e.
\begin{eqnarray}
iM=\frac{2(ad-bc)}{q}\,e
\end{eqnarray}
and
\begin{eqnarray}
iJ_{\mu\nu}=\frac{2(ac+bd)}{q\,\alpha_3}\,e_{\mu\nu}
\end{eqnarray}
Note that consistency now requires
\[
\frac{c^2+d^2}{\alpha_1}=\frac{a^2+b^2}{\alpha_2}=\frac{ac+bd}{\alpha_3}
\]
Next consider $\left[P_\mu,iM\right]=\alpha_3P_\mu-\alpha_1X_\mu$. We have
\begin{eqnarray}
\notag\left[P_\mu,iM\right]&=&\left[dee_\mu+ce_\mu,\frac{2(ad-bc)}{q}\,e\right]\\
\notag&=&\frac{2(ad-bc)}{q}\left[dee_\mu+ce_\mu,e\right]\\
\notag&=&\frac{2(ad-bc)}{q}(dee_\mu e+ce_\mu e-de^2e_\mu-cee_\mu)\\
\notag&=&\frac{2(ad-bc)}{q}(de_\mu-ce_\mu e-de_\mu-cee_\mu)\\
\notag&=&\frac{4(ad-bc)}{q}(de_\mu-cee_\mu)\\
\notag&=&\frac{4\Delta}{q}\left(\frac{d}{\Delta}(dX_\mu-bP_\mu)-\frac{c}{\Delta}(-cX_\mu+aP_\mu)\right)\\
\notag&=&\frac{4}{q}(d^2X_\mu-bdP_\mu+c^2X_\mu-acP_\mu)\\
&=&-\frac{4}{q}(ac+bd)P_\mu+\frac{4}{q}(c^2+d^2)X_\mu
\end{eqnarray}
For this to be equal to
\[
[P_{\mu},iM]=\alpha_3 P_\mu-\alpha_1 X_\mu
\]
we must have
\begin{eqnarray}
\alpha_3&=&-\frac{4}{q}(ac+bd)\\
\alpha_1&=&-\frac{4}{q}(c^2+d^2)
\end{eqnarray}
and the consistency conditions must now be extended to read
\begin{eqnarray}\label{consistency}
\frac{c^2+d^2}{\alpha_1}=\frac{a^2+b^2}{\alpha_2}=\frac{ac+bd}{\alpha_3}=-\frac{q}{4}
\end{eqnarray}
We also need to check $\left[X_\mu,iM\right]=\alpha_2P_\mu-\alpha_3X_\mu$. We have
\begin{eqnarray}
\notag\left[X_\mu,iM\right]&=&\left[ae_\mu+bee_\mu,\frac{2(ad-bc)}{q}\,e\right]\\
\notag&=&\frac{2(ad-bc)}{q}(ae_\mu e+bee_\mu e-aee_\mu-be^2e_\mu)\\
\notag&=&\frac{2(ad-bc)}{q}(-2aee_\mu+2be_\mu)\\
\notag&=&\frac{4\Delta}{q}(be_\mu-aee_\mu)\\
\notag&=&\frac{4\Delta}{q}\left(\frac{b}{\Delta}(dX_\mu-bP_\mu)-\frac{a}{\Delta}(-cX_\mu+aP_\mu)\right)\\
\notag&=&\frac{4}{q}(bdX_\mu-b^2P_\mu+acX_\mu-a^2P_\mu)\\
&=&-\frac{4}{q}(a^2+b^2)P_\mu+\frac{4}{q}(ac+bd)X_\mu
\end{eqnarray}

So we need
\begin{eqnarray}
\alpha_2&=&-\frac{4}{q}(a^2+b^2)\\
\alpha_3&=&-\frac{4}{q}(ac+bd)
\end{eqnarray}
which hold by the consistency equations.

To satisfy the consistency equation (\ref{consistency}), we may define
\[
a=\sqrt{-\frac{q\,\alpha_2}{4}}\;\cos\,\theta,\qquad
b=\sqrt{-\frac{q\,\alpha_2}{4}}\;\sin\,\theta
\]
 and
\[
c=\sqrt{-\frac{q\,\alpha_1}{4}}\;\sin\,\phi,\qquad
d=-\sqrt{-\frac{q\,\alpha_1}{4}}\;\cos\,\phi
\]
and since also $\displaystyle\frac{ac+bd}{\alpha_3}=-\frac{q}{4}$,
we obtain
\begin{eqnarray}
\sin(\theta-\phi)=\frac{\alpha_3}{\sqrt{\alpha_1\alpha_2}}
\end{eqnarray}

Comparing these results with those of Ahluwalia-Khalilova
\cite{Ahluwalia-Khalilova:2005jn}, we find that \footnote{In addition to the two conversions in (\ref{conversion}) above, there are two more conversions, one between $M$ and $F$ and one between $\alpha_3$ and $\beta$. A comparison reveals that consistency between \cite{Chryssomalakos:2004gk} and \cite{Ahluwalia-Khalilova:2005jn} requires that;
\begin{eqnarray}
\notag qM&=&F\\
\notag q\alpha_3&=&\beta\\
\notag q\alpha_2&=&\frac{1}{\ell_{c}^2}\\
\notag q\alpha_1&=&\ell_p^2
\end{eqnarray} }

\begin{equation}\label{conversion}
q\alpha_1=\frac{1}{\ell_{c}^2},\qquad q\alpha_2=\ell_{p}^2
\end{equation}

This in turn tells us that
\begin{eqnarray}
\alpha_3=\frac{\ell_{p}}{\ell_{c}}\sin(\varphi)
\end{eqnarray}
where $\varphi=\theta-\phi$.
In this section we have transformed the Clifford generators of the
representation to obtain a representation  where $\alpha_3$ is not equal to
zero. This is going in the opposite direction to Chryssomalakos and Okon \cite{Chryssomalakos:2004gk}
who begin with a representation where $\alpha_3$ is not necessarily
zero and then show that there always exists a representation in the
$\alpha_1$-$\alpha_2$ plane with $\alpha_3$ equal to zero by
performing a linear redefinition of the generators. The Clifford
algebra approach is thus consistent with \cite{Chryssomalakos:2004gk}.
\section{Physical Interpretation of Transformation}                  %
Chryssomalakos and Okon, \cite{Chryssomalakos:2004gk}, comment that physicists in particular may
frown upon the idea of working with arbitrary linear combinations of
momenta and positions. For this reason it is important that we
interpret what the transformation in the previous section might mean
physically.

The transformation
\begin{eqnarray}
\notag\omega_{1}e_\mu \mapsto
\sqrt{\frac{-q\alpha_2}{4}}\cos\theta e_\mu
+\sqrt{\frac{-q\alpha_2}{4}}\sin\theta ee_\mu
\end{eqnarray}
and
\begin{eqnarray}
\notag\omega_{2}\mathbf{ee}_\mu\mapsto
\sqrt{\frac{-q\alpha_1}{4}}\cos\phi ee_\mu+\sqrt{\frac{-q\alpha_1}{4}}\sin\phi e_\mu
\end{eqnarray}
is equivalent to adding some  momentum to the position vector and
vice versa. The magnitude of the vector is invariant.
The transformation looks like a rotation in the position-momentum
plane however $\omega_1 e_\mu$and $\omega_2
ee_{\mu}$ are rotated by different angles.

What is the physical interpretation of this transformation? In a
Newtonian mindset we consider time and space to be disjoint and one
can determine the absolute time and position of an event. Switching
to a relativistic mindset, we know that we can no longer treat space
and time separately but that in fact we have to consider them
together in what we call spacetime. It is no longer possible to
determine the time and position of an event absolutely.

Similarly, in a Newtonian mindset, we can treat  position and
momentum separately and thus the above transformation may seem
unphysical. In a quantum mechanical frame of mind however, we cannot
measure position $X_\mu$ and momentum $P_\mu$ with absolute
certainty. The more accurately we know one, the less accurately we
know the other as is described by the Heisenberg uncertainty
relationship $\Delta X_\mu \Delta P_\mu \geq \frac{\hbar}{2}$. This
suggests that we cannot just think about position or momentum
without considering the other. Furthermore, making a measurement of
the position of some particle will in itself affect the position.
The photon used to measure the particle's position will give the
particle some momentum. On the quantum scale therefore a linear
combination of position and momentum does make sense and in fact
treating position and momentum separately as in Newtonian physics
may no longer be desirable.

At the interface of GR and QM one combines quantum mechanics and relativity and
therefore should  consider spacetime-momentum instead of spacetime
alone \cite{Ahluwalia:1993dd}. 

\section{Acknowledgments}                  %
The authors of this paper wish to thank and acknowledge Dharamvir Ahluwalia-Khalilova, Benjamin Martin and Adam Gillard for the many discussions. In particular we wish to thank Dharamvir Ahluwalia-Khalilova for suggesting that at the interface of GR and QM one should consider spacetime-momentum instead of spacetime alone.

\begin{tabbing}
 {\em Email:}\ \  \= \kill
 {\em Email:}  \> {\tt ngg18@student.canterbury.ac.nz} (Gresnigt)\\
 \> {\tt P.Renaud@math.canterbury.ac.nz} (Renaud)\\
 \> {\tt phil.butler@canterbury.ac.nz} (Butler)\\
\end{tabbing}

\begin{thebibliography}{99}

\bibitem{Snyder:1946qz}
H.~S.~Snyder,
``Quantized space-time,''
Phys.\ Rev.\ {\bf 71} (1947) 38.


\bibitem{Yang:1947ud}
C.~N.~Yang,
``On quantized space-time,''
Phys.\ Rev.\ {\bf 72} (1947) 874.

\bibitem{VilelaMendes:1994zg}
  R.~Vilela Mendes,
  ``Deformations, stable theories and fundamental constants,''
  J.\ Phys.\ A {\bf 27} (1994) 8091.
  
\bibitem{Faddeev} L. D. Faddeev 1989, ``Mathematician's view on the development of physics, Frontiers in Physics: High technology and mathematics'' ed H. A. Cerdeira and S. Lundqvist (Singapore: Word Scientific, 1989) pp. 238-46
  
\bibitem{Chryssomalakos:2004gk}
  C.~Chryssomalakos and E.~Okon,
  ``Generalized quantum relativistic kinematics: A stability point of view,''
  Int.\ J.\ Mod.\ Phys.\ D {\bf 13} (2004) 2003
  [arXiv:hep-th/0410212].
  
\bibitem{Chryssomalakos:2004wc}
  C.~Chryssomalakos and E.~Okon,
  ``Linear Form of 3-scale Relativity Algebra and the Relevance of Stability,''
  Int.\ J.\ Mod.\ Phys.\ D {\bf 13} (2004) 1817
  [arXiv:hep-th/0407080].
  
\bibitem{Ahluwalia-Khalilova:2005jn}
  D.~V.~Ahluwalia-Khalilova,
   ``A freely falling frame at the interface of gravitational and quantum
  realms,''
  Class.\ Quant.\ Grav.\  {\bf 22} (2005) 1433
  [arXiv:hep-th/0503141].
  
\bibitem{Ahluwalia-Khalilova:2005km}
  D.~V.~Ahluwalia-Khalilova,
   ``Minimal spatio-temporal extent of events, neutrinos, and the  cosmological
  constant problem,''
  Int.\ J.\ Mod.\ Phys.\ D {\bf 14} (2005) 2151
  [arXiv:hep-th/0505124].
  
\bibitem{Ahluwalia:1993dd}
  D.~V.~Ahluwalia,
  ``Quantum Measurements, Gravitation, And Locality,''
  Phys.\ Lett.\ B {\bf 339} (1994) 301
  [arXiv:gr-qc/9308007].
  
\bibitem{NandR} A. Nijenhuis and R. W. Richardson 1967  ``Deformations of Lie algebra structures'' J. Math. Mech. $\mathbf{17}$ 89

\bibitem{Gerstenhaber} M. Gerstenhaber 1964 ``On the Deformation of Rings and Algebras'' Ann. Math. $\mathbf{79}$ 59-103
  
\bibitem{Khruschev:2002cq}
  V.~V.~Khruschev and A.~N.~Leznov,
   ``Relativistically invariant Lie algebras for kinematic observables in
  quantum space-time,''
  Grav.\ Cosmol.\  {\bf 9} (2003) 159
  [arXiv:hep-th/0207082].
  
\bibitem{dl}C. Doran and A. Lasenby 2003 ``Geometric Algebra for Physicists'' (Cambridge University Press)

\bibitem{pl}P. Lounesto 1997 ``Clifford Algebras and Spinors'',
  London Mathematical Society Lecture Note Series 239 (Cambridge
  University Press)
  
\bibitem{dh1}D. Hestenes 1966 ``Space--Time Algebra'' (Gordon and
  Breach, New York)
  
\bibitem{dh5}D. Hestenes 2003 ``Spacetime physics with geometric algebra'', \, Am.\ J. Phys. {\bf 71} (7), 691--714

\bibitem{wjpb}W. P. Joyce and P. H. Butler 2002 ``The Geometric
    Associative Algebras of Euclidean Space'' Adv.\ Appl.\ Clifford
  Alg.\ {\bf 12} (2), 195--233
  
\bibitem{pblm}P. Butler and L. McAven 1998 ``Space: The Anti-Euclidean Metric from the Structure of Rotations'', \, In Proceedings of the XII International Colloquium on Group Theoretical Methods in Physics, 494--498 (International Press)
  
\bibitem{Amelino-Camelia:2003bt}
  G.~Amelino-Camelia,
   ``Proposal of a second generation of quantum-gravity-motivated
   Lorentz-symmetry tests: Sensitivity to effects suppressed quadratically  by
  the Planck scale,''
  Int.\ J.\ Mod.\ Phys.\ D {\bf 12} (2003) 1633
  [arXiv:gr-qc/0305057].
  
\bibitem{Kowalski-Glikman:2004kp}
  J.~Kowalski-Glikman and L.~Smolin,
  ``Triply special relativity,''
  Phys.\ Rev.\ D {\bf 70} (2004) 065020
  [arXiv:hep-th/0406276].
  
\bibitem{Judes:2002bw}
S.~Judes and M.~Visser,
``Conservation Laws in Doubly Special Relativity,''
Phys.\ Rev.\ D {\bf 68} (2003) 045001

\bibitem{Grumiller:2003df}
D.~Grumiller, W.~Kummer and D.~V.~Vassilevich,
``A note on the triviality of kappa-deformations of gravity,''
Ukr.\ J.\ Phys.\ {\bf 48} (2003) 329
[arXiv:hep-th/0301061].

\bibitem{Schutzhold:2003yp}
R.~Schutzhold and W.~G.~Unruh,
``Problems of doubly special relativity with variable speed of light,''
JETP Lett.\ {\bf 78} (2003) 431
[Pisma Zh.\ Eksp.\ Teor.\ Fiz.\ {\bf 78} (2003) 899]
[arXiv:gr-qc/0308049].

\bibitem{Christian:2006ut}
  J.~Christian,
  ``Absolute being vs relative becoming,''
  arXiv:gr-qc/0610049.
  
\bibitem{Ahluwalia3} D. V. Ahluwalia-Khalilova 2006 {\it Private communications}
\end{thebibliography}
\end{document}